\documentclass[sigconf]{acmart}

\setcopyright{acmlicensed}
\copyrightyear{2026}
\acmYear{2026}
\acmDOI{XXXXXXX.XXXXXXX}
\acmConference[RecSys '26]{the Twentieth ACM Conference on Recommender Systems}{September 28–October 2, 2026}{Minneapolis, Minnesota, USA}
\acmISBN{978-1-4503-XXXX-X/2026/09}

\usepackage{amsmath}
\usepackage{amssymb}
\usepackage{booktabs}
\usepackage{multirow}
\usepackage{siunitx}
\usepackage{graphicx}   % 调整表格宽度
\usepackage{subcaption}
\usepackage{tcolorbox}
\usepackage{tabularx}  % 对应 \begin{tabularx} 和 {X} 参数
\tcbset{colback=white,colframe=black,boxrule=0.5pt}

\begin{document}

\title{Guess Where You Go: Generative Next Point-of-Interest Recommendation in Amap}

\author{Penglong Zhai}
\authornote{Contributed equally to this research.}
\affiliation{%
  \institution{AMAP, Alibaba Group}
  \city{Beijing}
  \country{China}
}
\email{zhaipenglong.zpl@alibaba-inc.com}

\author{Bowen Zheng}
\authornotemark[1]
\affiliation{
  \institution{AMAP, Alibaba Group}
  \city{Beijing}
  % \state{Ohio}
  \country{China}
}
\email{zbw405521@alibaba-inc.com}

\author{Jie Li}
\authornotemark[1]
\affiliation{%
  \institution{AMAP, Alibaba Group}
  \city{Beijing}
  \country{China}
}
\email{lj313796@alibaba-inc.com}

\author{Yifang Yuan}
\affiliation{%
  \institution{AMAP, Alibaba Group}
  \city{Beijing}
  \country{China}
}
\email{yuanyifang.yyf@alibaba-inc.com}

\author{Yue Liu}
\affiliation{%
  \institution{AMAP, Alibaba Group}
  \city{Beijing}
  \country{China}
}
\email{ly355576@alibaba-inc.com}

\author{Sicong Wang}
\affiliation{%
  \institution{AMAP, Alibaba Group}
  \city{Beijing}
  \country{China}
}
\email{wsc488938@alibaba-inc.com}

\author{Mingyang Yin}
\affiliation{%
  \institution{AMAP, Alibaba Group}
  \city{Beijing}
  \country{China}
}
\email{hengyang.ymy@amap.com}

\author{Tingting Hu}
\affiliation{%
  \institution{AMAP, Alibaba Group}
  \city{Beijing}
  \country{China}
}
\email{jingting.htt@alibaba-inc.com}

\author{Shuaijun Guo}
\affiliation{
  \institution{AMAP, Alibaba Group}
  \city{Beijing}
  % \state{Ohio}
  \country{China}
}
\email{guoshuaijun.gsj@alibaba-inc.com}

\author{Fanyi Di}
\authornotemark[2]
\affiliation{%
  \institution{AMAP, Alibaba Group}
  \city{Beijing}
  \country{China}
}
\email{difanyi.dfy@alibaba-inc.com}

\author{Xin Li}
\authornote{Corresponding Author.}
\affiliation{%
  \institution{AMAP, Alibaba Group}
  \city{Beijing}
  \country{China}
}
\email{beilai.bl@alibaba-inc.com}

\renewcommand{\shortauthors}{Zhai et al.}

\begin{abstract}

Generative retrieval enables recommender systems to retrieve items by generating compact item identifiers, but scaling it to industrial scenarios remains challenging due to redundant or colliding token assignments and insufficient integration of heterogeneous item signals. These challenges are particularly critical for next Point-of-Interest (POI) recommendation, where models must represent structured spatial entities, capture sequential mobility patterns, and produce predictions consistent with real user behavior.
We propose \textbf{Gwhere}, an end-to-end industrial framework that integrates semantic identifier (SID) generation with LLM-based generative next POI recommendation. Gwhere first learns discriminative POI SIDs through a contrastive residual-quantization tokenizer that aligns textual, visual, spatial, and collaborative signals. Based on these SIDs, Gwhere adapts LLMs to mobility scenarios via continued pretraining on enriched spatio-temporal corpora, supervised fine-tuning, and Exposure-Aware Kahneman-Tversky Optimization (EAKTO), a reinforcement learning objective for behavioral preference alignment.
Experiments on public datasets and Amap's large-scale industrial dataset demonstrate the effectiveness of Gwhere. The system has been deployed in Amap's homepage service under high-concurrency and low-latency constraints. Long-term online A/B tests show improvements of 5.83\% in P-CTR and 6.20\% in U-CTR over the production baseline. The implementation is publicly available at https://github.com/alibaba/SimCIT.
\end{abstract}




\maketitle

\section{Introduction}

Generative retrieval has emerged as a promising paradigm for recommendation by reformulating item retrieval as a sequence generation problem \cite{deldjoo2024review,Vaswani2017,Devlin2019,Rajput2023}. A key component is item tokenization, which maps large item-ID spaces into compact semantic identifiers (SIDs) and enables efficient autoregressive decoding. However, predominant VAE-based tokenizers, such as RQ-VAE, are mainly optimized by reconstruction loss. This objective does not directly enforce inter-item separability, which is crucial for generative retrieval, and may produce ambiguous or collapsed token assignments for semantically similar and long-tail items. Moreover, existing tokenization methods often integrate textual, visual, and spatial signals through shallow fusion mechanisms, limiting their ability to capture hierarchical multi-modal semantics under industrial latency and memory constraints.

We focus on next Point-of-Interest (POI) prediction, a fundamental location-based service task that predicts users' future locations from historical check-ins and contextual signals. Although Large Language Models (LLMs) provide strong sequence modeling and semantic reasoning capabilities, directly applying them to POI recommendation remains challenging. First, raw POI identifiers are not semantically grounded in LLM vocabularies. Second, LLMs struggle to encode complex user-POI collaborative signals. Third, their predictions may violate human spatial cognition, such as distance awareness, familiarity, and common-sense mobility preferences.

To address these challenges, we propose Gwhere, an end-to-end generative framework for next POI prediction. Gwhere first learns compact multi-modal SIDs through a fully contrastive tokenization framework with hierarchical identifier learning, improving item discriminability while integrating heterogeneous POI information. It then adapts LLMs to the mobility domain through continued pretraining on large-scale spatio-temporal POI sequences and a multi-stage fine-tuning pipeline, including Supervised Fine-Tuning (SFT) and Exposure-Aware Kahneman-Tversky Optimization (EAKTO), to improve both predictive accuracy and cognitive plausibility.

Our main contributions are summarized as follows:
\begin{itemize}
\item We propose a fully contrastive semantic tokenization framework for generative recommendation. By combining learnable residual quantization with hierarchical identifier learning, it integrates heterogeneous POI signals, including textual, visual, spatial, and collaborative information, and improves token discriminability over VAE-based tokenizers.

\item We present Gwhere, an end-to-end LLM-based framework for cognitively aligned next POI prediction. Gwhere unifies multi-modal semantic identifiers with generative mobility modeling, and adapts LLMs to the mobility domain through continued pretraining, supervised fine-tuning, and Exposure-Aware Kahneman-Tversky Optimization (EAKTO).

\item We validate Gwhere through extensive offline experiments and large-scale online deployment in Amap's production environment. Long-term A/B tests show consistent improvements of 5.83\% in P-CTR and 6.20\% in U-CTR over the production baseline.
\end{itemize}

The remainder of this paper is organized as follows. Section~\ref{sec:related} reviews related work on
  generative retrieval, semantic tokenization, and next POI recommendation. Section~\ref{sec:method} introduces
  our contrastive multi-modal item tokenization framework and presents the Gwhere framework for generative next
  POI prediction based on SIDs. Section~\ref{sec:exp} reports offline and online experimental results.
  Section~\ref{sec:conclusion} concludes the paper and discusses future directions.

\section{Related Work}
\label{sec:related}
\subsection{Next POI Prediction}
Next POI recommendation extends beyond conventional item recommendation by incorporating inherent geographical and temporal characteristics. Early methods relied on user check-in histories using Markov chains and RNNs to capture sequential behaviors. Spatial-temporal modeling has been extensively studied with works like ST-GRAT \cite{Park_2020}, STAN \cite{chen2023stan}, and Sun et al. \cite{sun2021periodicmove}.
Recent LLM-based approaches have focused on addressing the semantic gap in POI recommendation. For example, GNPR-SID \cite{wang2025generative} proposed semantic POI IDs using residual quantized variational autoencoders, and LLM4POI \cite{zhang2024llm4poi} leveraged pretrained LLMs to preserve heterogeneous data in original format.

\subsection{LLMs for Recommendation}
The integration of Large Language Models (LLMs) into recommendation systems has emerged as a transformative research direction. P5 \cite{geng2023recommendationlanguageprocessingrlp} pioneered the "Pretrain, Personalized Prompt, and Predict" paradigm, establishing the foundation for treating recommendation as a language processing task. Building upon this, addressed the knowledge gap between LLMs' capabilities and recommendation-specific requirements by incorporating collaborative filtering operations like Masked Item Modeling (MIM) and Bayesian Personalized Ranking (BPR) through natural language simulation. More recently, OneRec \cite{deng2025onerecunifyingretrieverank, zhou2025onerectechnicalreport} proposed a revolutionary end-to-end framework that replaces traditional cascaded architectures.

\subsection{Semantic Indexing and Tokenization}
Beyond the core generation framework, how to effectively index items has gained significant attention. Straightforward indexing methods like random or title-based indexing are simple, but they do not scale to industrial-size retrieval systems and lack rich semantic information. To tackle this, recent research \cite{qu2024tokenreclearningtokenizeid, 10.1145/3627673.3679569, 10.1145/3640457.3688178, 10.1145/3640457.3688190, 10.5555/3600270.3601857} has gravitated towards semantic indexing techniques, which aim to categorize items based on their inherent content information. Within this area, two primary techniques stand out: vector quantization and hierarchical k-means clustering. For example, TIGER\cite{Rajput2023} and LC-Rec\cite{zheng2024adaptinglargelanguagemodels} use residual quantization (RQ-VAE) on textual embeddings derived from item titles and descriptions for tokenization. On the other hand, Recforest\cite{10.5555/3600270.3603090} and EAGER\cite{Wang2024EAGERTG} employ hierarchical k-means clustering on item textual embeddings to create cluster indexes as tokens. However, few studies have addressed how to use one framework to effectively fuse many different sources and multi-modal information to generate effective tokens.

\section{Methodology}
\label{sec:method}
This section elaborates on the integrated framework (Gwhere) for next POI prediction, which unifies Semantic IDs (SIDs) generation and spatio-temporal LLM post-training. We first formalize the problem, then detail the motivation and design of each core component, and finally describe the inference optimization for industrial deployment.

\subsection{Problem Formulation}
% \subsubsection{ Next POI Prediction Task}
Given a set of POIs $\mathcal{I} = \{p_1, p_2, \ldots, p_N\}$ and users $\mathcal{U} = \{u_1, u_2, \ldots, u_M\}$, each POI $p_i$ is represented as $(lon, lat, \mathcal{C})$, where $lon$ and $lat$ are geographical coordinates, and $\mathcal{C}$ denotes multi-modal side information (e.g., textual descriptions, images, category tags). Each user $u$ has a historical check-in sequence $h_u = \{(p_{u,1}, t_{u,1}), ..., (p_{u,m}, t_{u,m})\}$, where $p_{u,i} \in \mathcal{I}$ is the $i$-th interacted POI and $t_{u,i}$ is the corresponding timestamp. Given $h_u$, the user's current location $l_u$, and contextual situation $s_u$ (e.g., weather, holidays, user profile), the goal is to predict the next POI $p_{u,m+1}$ that user $u$ will visit at time $t_{u,m+1}$.

% \subsubsection{Generative POI Prediction Paradigm}
In the generative paradigm, each POI $p_i$ is mapped to a discrete SIDs $sid(p_i) = (c_1^i, ..., c_L^i)$ through item tokenization, where $L$ is the length of the SIDs tuple and $c_k^i$ denotes the $k$-th codeword from the codebook. The historical check-in sequence $h_u$ is converted into a sequence of SIDs $S_u = [sid(p_{u,1}), sid(p_{u,2}), ..., sid(p_{u,m})]$. The next POI prediction is reformulated as autoregressively generating $sid(p_{u,m+1})$ from $S_u$, $l_u$, and $s_u$, formally defined as:
\vspace{-2.7pt}
\begin{equation*}
p(p_{u,m+1} | h_u, l_u, s_u) = \prod_{k=1}^L p(c_k^{u,m+1} | S_u, l_u, s_u, c_1^{u,m+1}, ..., c_{k-1}^{u,m+1})
\end{equation*}
After generating the target SIDs, we map them back to the corresponding POI via a pre-built SID-to-POI lookup table to complete the recommendation.

\subsection{Contrastive Item Tokenization for SIDs}
Item tokenization is a critical prerequisite for generative POI recommendation, but existing methods face two key limitations: (1) Predominant reconstruction-based quantization strategies (e.g., RQ-VAE) optimize via MSE loss, which conflicts with the core objective of generative retrieval—robust differentiation among items, leading to representation collapse in long-tail POI scenarios. (2) Multi-modal information is fused via primitive mechanisms (e.g., input-layer concatenation), failing to capture hierarchical semantic dependencies. To address these, we propose a contrastive item tokenization framework that eliminates reconstruction loss and leverages contrastive learning to enhance SID discriminability and multi-modal alignment.

\subsubsection{Multi-modal Feature Extraction and Fusion}
We extract four types of modal features for each POI: \textbf{textual} embeddings $z_\text{text}$ via a pre-trained language model (e.g., BERT \cite{Devlin2019}); \textbf{visual} embeddings $z_\text{img}$ via a pre-trained Vision Transformer \cite{dosovitskiy2021imageworth16x16words}; \textbf{spatial} embeddings $z_\text{spatial}$ via a GraphSAGE \cite{10.5555/3294771.3294869} encoder over a spatial graph with Haversine-distance edges; and \textbf{collaborative} embeddings $z_{cf}$ via alternating least squares (ALS) on the user-POI interaction matrix.

Different modalities contribute unequally to POI tokenization. Each modality embedding is first projected into a shared $d$-dimensional space, and we adopt an attention-based fusion strategy to adaptively weight each modality:
\[
a_m = \frac{\exp(q^\top z_m)}{\sum_{m' \in \mathcal{M}} \exp(q^\top z_{m'})},
\qquad
z = \sum_{m \in \mathcal{M}} a_m z_m .
\]
where $q$ is a learnable attention vector and $\mathcal{M}$ is the set of available modalities.

\subsubsection{Soft Residual Quantization with Contrastive Learning}
We employ residual quantization to discretize the fused embedding into SIDs, replacing reconstruction loss with contrastive learning for better discriminability.

\paragraph{Differentiable Quantization}
We initialize $L$ learnable codebooks $\mathcal{C}_l = \{e_k^l | k=1,...,K\}$. Starting from residual $r_0=z$, at each layer $l$ we compute a Gumbel-softmax soft assignment $c_k^l \propto \exp\big((-\| r_{l-1}-e_k^l \|_2^2 + \epsilon_k^l)/\alpha\big)$ with $\epsilon_k^l \sim \mathrm{Gumbel}(0,1)$ and annealed temperature $\alpha$, and update the residual via $r_l = r_{l-1} - \sum_{k} c_k^l e_k^l$. The quantized representation is
\[
\hat{z} = \sum_{l=1}^L \sum_{k=1}^K c_k^l e_k^l .
\]

\paragraph{Contrastive Loss Optimization}
To ensure SIDs of semantically similar POIs are close while dissimilar ones are far apart, we optimize the NT-Xent loss \cite{Chen2020SimCLR} between $\hat{z}$ and all modal representations:
\[
\mathcal{L} = -\sum_{m=1}^{|\mathcal{M}|} \log \frac{\exp(\hat{h} \cdot h_m^+ / \tau)}{\sum_{h^- \in \mathcal{B}} \exp(\hat{h} \cdot h^- / \tau)}
\]
where $\hat{h} = g(\hat{z})$ and $h_m^+ = g(z_m)$ are projections via a 2-layer MLP $g$, $\tau=0.1$ is the temperature hyperparameter, and $\mathcal{B}$ denotes negative samples within a batch.
\subsection{Spatio-Temporal LLM Post-Training}
% \subsubsection{3.3.1 Motivation}
While SIDs reduce token space and preserve semantics, directly applying pre-trained LLMs to POI generation faces three critical challenges: (1) LLMs are pretrained on unstructured text and lack native understanding of structured geographical entities (e.g., Geohash) and sequential mobility patterns (e.g., commute rhythms). (2) Recommendations often violate human cognitive preferences (e.g., outdoor POIs on rainy days), eroding user trust and retention. (3) POIs are indexed by discrete identifiers outside the LLM vocabulary, impeding semantic relationship capture. To address these, we design a spatio-temporal LLM post-training framework that adapts LLMs to the mobility domain and aligns outputs with human preference.

\subsubsection{Spatio-Temporal Knowledge Acquisition via Continued Pretraining}
Continued pretraining is a critical step to adapt base LLMs (e.g., Qwen2.5 \cite{qwen2025qwen25technicalreport}) to POI recommendation, addressing three core limitations of vanilla LLMs: (1) SIDs and Geohash tokens are out-of-vocabulary (OOV) for pre-trained LLMs, requiring explicit semantic alignment between tokens and real-world POI attributes; (2) LLMs lack the ability to model spatio-temporal user behavior patterns (e.g., weekly commute, weekend leisure) that are fundamental to next POI prediction; (3) The gap between general language understanding and recommendation-specific knowledge (e.g., collaborative signals, POI category correlations) needs to be bridged. Below, we detail the design of pretraining corpora and the objectives of continued pretraining.

We construct two large-scale, domain-specific corpora to cover both semantic alignment and behavior pattern learning, with all data derived from desensitized Amap user trajectories and POI metadata. Examples are shown in Table~\ref{tab:pretrain-corpus}.

% \begin{table}[h]
% \centering
% \caption{Examples of recommendation corpora collected for continuous pretraining.}
% \label{tab:pretrain-corpus}
% \begin{tabularx}{\linewidth}{X}
% \toprule
% \textbf{SID-Location-Description Corpus} \\
% \midrule
% The Temple of Heaven Park is a scenic spot in Beijing, located at \texttt{<wm6j0>} (Geohash). Its SIDs is \texttt{<a\_82><b\_59><c\_191>}. \\
% \midrule
% \textbf{Users' Behavior Sequence Corpus} \\
% \midrule
% User \texttt{uid\_256}'s check-in history: 7 AM June 10 (weekday) went \texttt{<a\_124><b\_192><c\_41>} (company); 6 PM June 10 searched \texttt{<a\_82><b\_59><c\_191>} (restaurant); 9 PM June 10 navigate to \texttt{<a\_124><b\_192><c\_41>} (home). \\
% \bottomrule
% \end{tabularx}
% \end{table}
 \begin{table}[h]
  \centering
  \caption{Examples of recommendation corpora collected for continuous pretraining.}
  \label{tab:pretrain-corpus}
  \begin{tabularx}{\linewidth}{X}
  \toprule
  \textbf{SID-Location-Description Corpus} \\
  \midrule
  The Temple of Heaven Park is a scenic spot in Beijing, located at \texttt{<wm6j0>} (Geohash). Its SIDs is
  \texttt{<a\_82><b\_59><c\_191>}. \\
  \midrule
  Starbucks (Wangjing SOHO Branch) is a coffee shop in Beijing, located at \texttt{<wx4g0>} (Geohash). Its SIDs is
  \texttt{<a\_34><b\_127><c\_88>}. \\
  \midrule
  \textbf{Users' Behavior Sequence Corpus} \\
  \midrule
  User \texttt{uid\_256}'s check-in history: 7 AM June 10 (weekday) went \texttt{<a\_124><b\_192><c\_41>} (company); 6
  PM June 10 searched \texttt{<a\_82><b\_59><c\_191>} (restaurant); 9 PM June 10 navigate to
  \texttt{<a\_124><b\_192><c\_41>} (home). \\
  \midrule
  User \texttt{uid\_512}'s check-in history: 10 AM Saturday (weekend) drove to \texttt{<a\_12><b\_28><c\_140>} (park); 2
   PM Saturday walked to \texttt{<a\_67><b\_45><c\_88>} (cafe). \\
  \bottomrule
  \end{tabularx}
  \end{table}

\paragraph{SID-Location-Description Corpus}
This corpus aims to establish bidirectional semantic alignment between SIDs, POI geographical locations, and textual descriptions—solving the OOV token problem and enabling LLMs to ``understand'' what a SID represents. Each entry is a triplet of (SID, Geohash, Text Description), where \textbf{Geohash} is a 6-character string encoding the POI's precise geographical coordinates, enabling the model to learn spatial proximity (e.g., SIDs with similar Geohashes are geographically close).

\paragraph{Spatio-Temporal Behavior Sequence Corpus}
This corpus focuses on teaching LLMs to model user mobility patterns by learning sequential dependencies between SIDs, timestamps, and contextual attributes. Each entry is a time-ordered sequence of user check-ins, converted into a natural language narrative with SIDs. Key design elements include: (1). Check-in Granularity: We retain the most recent check-ins per user (balancing sequence length and computational efficiency), each annotated with timestamp and action type; (2). Contextual Annotations: We add situational context (e.g., ``workday morning'') and user profile hints (e.g., ``office worker'') to enable pattern learning.

The continued pretraining optimizes two complementary objectives to achieve the dual goals of semantic alignment and behavior modeling:
1. {Masked SID Prediction}: For the behavior sequence corpus, the model learns to predict masked SIDs based on surrounding check-ins, timestamps, and context—training it to capture sequential dependencies and spatial correlations (e.g., restaurants near company are more likely to be visited at lunchtime).
2. {SID-Attribute Alignment}: For the alignment corpus, we randomly shuffle SIDs of entries and train the model to distinguish matched (SID $\leftrightarrow$ Geohash $\leftrightarrow$ Description) from mismatched triplets—reinforcing semantic consistency.

Through these objectives, the LLM achieves three key capabilities after pretraining: (1) It can map SIDs to real-world POI attributes (location, category, function); (2) It understands spatial relationships (e.g., geographically close SIDs are semantically related); (3) It learns common user mobility patterns (e.g., time-dependent POI transitions).

\subsubsection{Supervised Fine-Tuning (SFT)}
The SFT phase aims to teach the model explicit recommendation logic and preliminary user preference alignment, laying the foundation for subsequent preference refinement.

We construct the SFT dataset using high-quality user interaction data and multi-level user profiles (Static Profile, Long-term Preference Profile, Periodic Demand Profile), formatting each sample as a prompt-response pair:
\begin{tcolorbox}
\textbf{Prompt:} \\
User profile: \{static profile\}; \{long-term preference\}; \{periodic demand\} (recurring patterns). Historical check-ins: \{SID sequence of recent check-ins\}. Current context: location=\{l\}, time=\{t\} situation=\{s\} (weather/traffic). Please recommend the next POI that the user is likely to prefer.\\
\textbf{Response:} \\
SIDs of the POI that the user actually clicked or interacted with (reflecting explicit preference), e.g., \texttt{<a\_112><b\_32><c\_5>}.
\end{tcolorbox}
We fine-tune the model using cross-entropy loss, optimizing it to generate POI SIDs that align with user historical preferences and contextual constraints. After SFT, the model can generate semantically reasonable recommendations and adhere to the required output format, but still needs to be optimized for real-world preference ranking and noise-resistant capability.

\subsection{Preference Alignment from Implicit User Feedback}
Following Supervised Fine-Tuning stage, the model has mastered spatio-temporal patterns, SID-semantic alignment, and basic recommendation knowledge. 
However, SFT relies on explicit interaction labels but fails to model the relative preference relationships between POIs and is insufficient in capturing dynamic user preferences in real scenarios (e.g., transient interests, context-dependent choices). 
Standard preference aligning methods in LLMs (DPO \cite{rafailov2023direct}, GRPO\cite{grpo}) require paired comparison data \((x, y_w, y_l)\), which is not naturally available in large-scale recommendation systems.
In contrast, KTO \cite{KTO}
operates on simpler binary signals (desirable / undesirable), aligning with the natural structure of exposure logs (clicks vs. non-clicks). This enables scalable, low-cost preference alignment, critical for improving ranking performance in industrial settings.
To further enhance its ability to capture fine-grained user preferences, we propose Exposure-Aware Kahneman-Tversky Optimization (EAKTO), leveraging real-world exposure logs to refine preference modeling and ranking performance.

% \subsubsection{Data Construction for KTO}
% We leverage Amap's online exposure logs to construct the KTO dataset:
% (1). For each request context \(x\) (historical SID sequence, user profile, current location, time, and situation), collect the exposed POI slate.
% (2). Treat clicked POIs \(y^+\) as desirable examples (\(s=+1\)) and exposed-but-unclicked POIs \(y^- \in \mathcal{E}(x)\) as exposure-based negatives (E-Neg, \(s=-1\)).Form triples \((x, y, s)\) to model binary preference signals, avoiding manual annotation costs.
% This data construction method is scalable and cost-effective, as it reuses existing exposure logs without additional annotation, and the binary signals directly reflect user preference feedback for ranking optimization—addressing the data scarcity issue of paired comparison methods.

% % \subsubsection{ Challenges of Standard KTO in Recommendation}
% Standard KTO faces two key issues in next POI prediction:
% 1. \textbf{Reward Drift}: Prolonged training drives rewards of both desirable/undesirable samples negative, undermining preference for clicked items.
% 2. \textbf{Noisy Negatives}: Non-clicks often stem from transient factors (inattention, latency) rather than true dis-preference, leading to over-conservatism.

% \subsubsection{EAKTO}
However, standard KTO faces two key issues in recommendation scenario:
\begin{itemize}
    \item \textbf{Reward Drift}: Prolonged training drives rewards of both desirable/undesirable samples negative, undermining preference for clicked items.
    \item \textbf{Noisy Negatives}: Non-clicks often stem from transient factors (inattention, latency) rather than true dis-preference, leading to over-conservatism.
\end{itemize}
To address these challenges, we propose Exposure-Aware Kahneman-Tversky Optimization (EAKTO), with two core refinements:
1. \textbf{Positive Reward Anchoring}: Enforce \(r_\theta(x,y) > 0\) for desirable samples to prevent reward drift.
2. \textbf{Conditional Negative Update}: Compute loss for negatives only when \(r_\theta(x,y) > 0\) (erroneous positive assignment), suppressing noisy negative impacts.

Specifically, the EAKTO loss is formulated as:
\begin{equation}
  \begin{aligned}
    \mathcal{L}_{\text{EAKTO}}(\pi_\theta, \pi_{\text{ref}}) = 
    \mathbb{E}_{x,y \sim \mathcal{D}} \Big[ &\lambda_y - v(x,y) \\
    &+ \alpha \cdot \mathbb{I}_{y \in \mathcal{Y}_{\text{D}}} \cdot \sigma_{\text{D}}(x,y) \Big]
  \end{aligned}
  \label{eq:eakto}
\end{equation}
where \(r_{\theta}(x, y) = \log \frac{\pi_{\theta}(y|x)}{\pi_{\text{ref}}(y|x)}\) (reward of output \(y\) relative to SFT reference model \(\pi_{\text{ref}}\)), \(z_0 = \text{KL}(\pi_{\theta}(y'|x) \parallel \pi_{\text{ref}}(y'|x))\) (regularization for update magnitude),
and \(v(x,y)\) (preference alignment term, differentiated for \(\mathcal{Y}_{\text{D}}\) (desirable) and \(\mathcal{Y}_{\text{U}}\) (undesirable)):
  \[
  v(x,y) = 
  \begin{cases} 
    \lambda_D \cdot \sigma(\beta (r_\theta(x,y) - z_0)) & y \in \mathcal{Y}_{\text{D}} \\
    \lambda_U \cdot \mathbb{I}_{\{r_\theta(x,y) > 0\}} \cdot \sigma(\beta (z_0 - r_\theta(x,y))) & y \in \mathcal{Y}_{\text{U}}
  \end{cases}
  \]
where \(\sigma_{\text{D}}(x,y) = \max(0, \log \frac{\pi_{\text{ref}}(y|x)}{\pi_\theta(y|x)})\) (noisy click regularization),
The hyperparameters are set as $\lambda_D=1.0$, $\lambda_U=1.0$, $\beta=0.1$, $\alpha=0.3$.

% \subsubsection{3.3.3.6 Key Contributions}
% To our knowledge, this is the first use of KTO for learning recommendation preferences from exposure logs. 
Empirically, EAKTO effectively addresses reward drift and noisy negatives, enhancing the model’s ability to capture true user preferences and improve ranking performance. In experiments, we compare SFT-only, SFT+GRPO, SFT+DPO, and SFT+EAKTO to isolate the preference objective’s impact.

\begin{figure*}
    \centering
    \includegraphics[width=\linewidth]{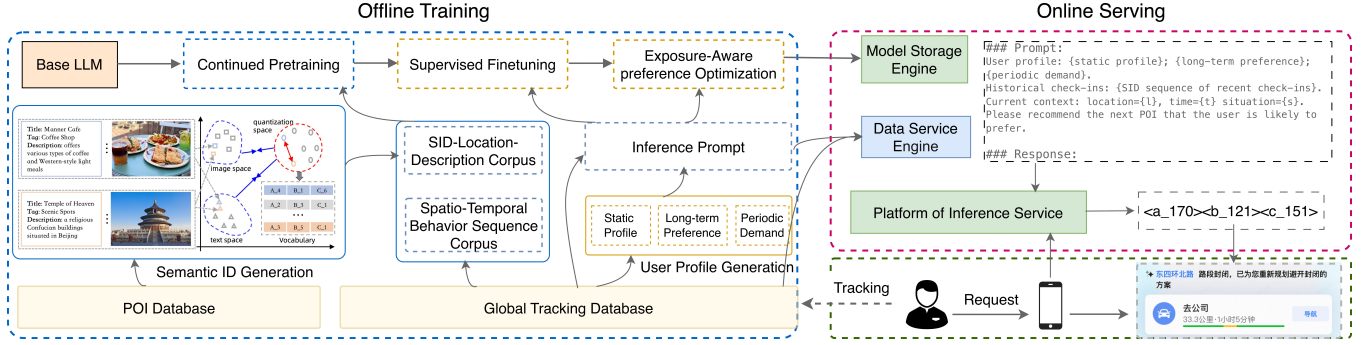}
    \caption{Offline Training and Online Serving Procedure of proposed framework Gwhere. Offline Training:(1) POIs are tokenized as Semantic IDs (SIDs); (2) a base LLM is continually pretrained on the collected corpus. (3) The LLM is further fine-tuned and aligned the with inference prompt to finish this procedure. Online Serving: Gwhere is stored in a model engine and invoked through a data-service engine that converts real-time context—user profile, recent check-ins, current time, location, and weather—into an instruction prompt. The model predicts the next POI in SID form, which is mapped back to a readable location and returned to the mobile client.}
    \label{fig:framework_gwhere}
\end{figure*}

\section{Experiments}
\label{sec:exp}
To validate the effectiveness of the proposed frameworks for next POI recommendation, we conduct extensive offline experiments on public Foursquare datasets and large-scale industrial Amap datasets, followed by online A/B tests on Amap’s production environment. The experiments aim to verify the model’s performance in prediction accuracy, user preference capture, ranking capability, and industrial applicability, with detailed settings and results derived from the original works.

\subsection{Experimental Settings}
% \subsubsection{Scenario} Our method is performed on \textbf{AMAP}, a prominent navigation and mapping platform hosting billions of users and POIs. We focus on the "\textbf{Guess Where You Go}" card presented on the homepage of the AMAP App, aiming to forecast the next subsequent POI using a user’s previous check-ins and profile information. 
% The objective of this scenario is to predict the next POI the users are most likely to visit. By displaying the POIs on the homepage, we aim to reduce the users' search effort, thereby enhancing user experience and long-term retention. 

% The current deployed framework, which serves as our primary baseline, employs a traditional cascaded ranking system for recommendation. As illustrated in Figure \ref{fig:cascade}, the cascaded ranking system includes three stages: (1) Retrieval stage to recall a coarse-grained POI corpus based on the user's preference, historical check-ins, and location \cite{10.1145/3357384.3357814,yang2020large}; (2) Ranking stage to obtain top 10 POIs the users most likely visit with a ranking model based on behavior sequence transformer \cite{10.1145/3326937.3341261}; and (3) Cognitive Re-ranking stage with distance, quality, and exposure filters to yield the most cognitively plausible destination.

% \begin{figure}[h!]
%     \centering
%     \includegraphics[width=\linewidth]{figs/cascade.jpg}
%     \caption{Cascade ranking system of "Guess Where You Go" deployed on AMAP, which includes three stages from the left to the right: Retrieval, Ranking, and Cognitive Re-ranking.}
%     \label{fig:cascade}
% \end{figure}

\subsubsection{Datasets}
\label{sec:dataset}
To validate our model's performance in a real-world application, we constructed a large-scale proprietary dataset by sampling desensitized users from Amap. For a fair comparison against existing baselines, we also evaluate on three real-world datasets: Foursquare-NYC \cite{6844862}, Foursquare-TKY \cite{6844862}, and Gowalla-CA \cite{10.1145/2020408.2020579}.

The pre-processing follows the settings in \cite{wang2025generative}: we remove POIs with fewer than 10 interactions and users with fewer than 10 check-ins, then sort by time with 80\%/10\%/10\% train/validation/test splits. Users and POIs absent from the training set are removed from the test set. Check-in records are grouped by user and ordered chronologically; for each user, the last visited POI is held out as the ground truth, while the preceding visits form the input sequence. Detailed statistics are provided in Table \ref{label:stat}.

\begin{table}[h!]
\centering
\setlength{\tabcolsep}{5.pt} % Adjusts column separation
\renewcommand{\arraystretch}{1.2} % Adjusts row separation
\caption{Statistics of the preprocessed datasets. Avg.len denotes the average length of item sequences.}
\label{label:stat}
\begin{tabular}{lccccc}
\toprule[1pt]
\textbf{Dataset} & \textbf{\#Users} & \textbf{\#Items} & \textbf{\#Inter.} & \textbf{Avg.len} & \textbf{Sparsity} \\
\midrule
% Pre-training & 999,334 & 344,412 & 8,609,909 & 8.62 & 99.997\% \\
NYC & 1,083 & 5,135 & 104,074 & 136 & 98.1013\%   \\
TKY & 2,293 & 7,873 & 361,430 & 195 & 97.9798\%  \\
CA & 6,592 & 14,027 & 349,375 & 53 & 99.9996\%  \\
AMAP  & 10,000k & 83,100k & 380,000k & 38 & 99.9999\%  \\
\bottomrule[1pt]
\end{tabular}
\end{table}

\subsubsection{Implementation Details}
For the SID tokenizer, we construct four modality-specific views for each POI: textual titles and descriptions embedded by a pretrained language encoder; collaborative signals from alternating least squares on the user--item interaction matrix; POI images encoded by a pretrained vision transformer \cite{dosovitskiy2021imageworth16x16words} when available; and spatial embeddings derived via a GraphSAGE-style encoder over a spatial graph built from geographic distance and mobility flows. 
% The four views are fused by a lightweight multi-layer perceptron with attention/gating to produce a unified continuous representation for each POI.

This representation is passed through a residual quantizer with multiple codebooks. In our default setup, a 96-dimensional latent vector is quantized by a 3-layer codebook stack with tens of codewords each (e.g., 48 for public datasets, larger for AMAP), so each POI is represented by a short tuple of codes. 
% The contrastive loss temperature and soft assignment schedules largely follow the SimCIT design (e.g., a small temperature such as 0.1 with gradual annealing). The tokenizer is trained with Adam using a learning rate around $10^{-4}$ and batch size around 256.
% For the LLM recommender, we adopt a Transformer-based decoder and reuse the same SID vocabulary across continued pretraining, supervised fine-tuning, and preference optimization. Unless otherwise stated, all variants in the ablation studies use identical architectures and only differ in whether they include SID tokenization, continued pretraining, SFT, and the specific preference objective (DPO, GRPO, or KTO).
% The quantization module consists of 3 codebook layers and 

For the generative recommendation module, we employ Qwen \cite{qwen2025qwen25technicalreport} as the base model, with a constant learning rate of 1e-5 and a 20-step warm-up. The model is trained on 200$\times$ NVIDIA H20 GPUs, with batch size 16 per GPU, gradient accumulation 8, and sequence length 4096. Each input consists of the user's most recent 50 check-in records.

\subsubsection{Baseline Methods}
Following the settings of \cite{wang2025generative,zhang2024llm4poi}, we compare Gwhere against five categories of state-of-the-art next POI recommendation baselines:
(1) Traditional: PRME \cite{10.5555/2832415.2832536} (metric embedding) and PLSPL \cite{9117156} (personalized sequential).
(2) Transformer-based: STAN \cite{chen2023stan} (spatio-temporal attention) and GETNext \cite{10.1145/3477495.3531983} (Transformer-based next POI recommendation).
(3) GCN-based: STHGCN \cite{10.1145/3539618.3591770} (spatio-temporal hierarchical graph).
(4) Time-aware: TPG \cite{10.1145/3583780.3615083} (timestamp-guided) and ROTAN \cite{10.1145/3637528.3671809} (time-aware POI recommendation).
(5) LLM-based: LLM4POI \cite{zhang2024llm4poi} and GNPR-SID \cite{wang2025generative}, a state-of-the-art generative framework with semantic codes, closely related to our work.

% INS and BEA are used for next-item prediction in e-commerce, while NYC, TKY, and AMAP are used for next POI prediction. AMAP is our primary focus for the industrial study, and the public datasets are used to verify that the learned semantic IDs and the unified pipeline generalize across domains. Table~\ref{tab:dataset_stats} summarizes basic statistics of these datasets; numerical values are omitted and can be filled in once the final data snapshot is fixed.

% \begin{table}[t]
%   \centering
%   \caption{Statistics of the datasets used in our experiments. ``Avg. length'' denotes the average length of user sequences.}
%   \label{tab:dataset_stats}
%   \begin{tabular}{lccccc}
%     \toprule
%     Dataset & \#Users & \#Items & \#Inter. & Avg.len & Sparsity \\
%     \midrule
%     INS & 57,439 & 24,587 & 511,836 & 8.91 & 99.9640\% \\
%     BEA & 50,985 & 25,848 & 412,947 & 8.10 & 99.9690\% \\
%     NYC & 1,075 & 5,099 & 104,074 & 96.8 & 98.1013\% \\
%     TKY & 2,281 & 7,844 & 361,430 & 158.4 & 97.9798\% \\
%     AMAP & 7,684k & 6,158k & 172,100k & 22.39 & 99.9996\% \\
%     \bottomrule
%   \end{tabular}
% \end{table}

\subsubsection{Our Models} We consider four versions of our model, Gwhere-\{0.5B, 1.5B, 3B, 7B\}, each trained with the corresponding Qwen2.5 \cite{qwen2025qwen25technicalreport} as the base LLM. Unless stated otherwise, "our model" refers to Gwhere-0.5B.

\begin{table}
  \caption{Comparison results of different versions of Gwhere on AMAP dataset with the Cascade Ranking baseline system.}
  \label{tab:beijing}
  \setlength{\tabcolsep}{2.5pt}
  \begin{tabular}{cccccc}
    \toprule
    &\small Baseline &\small Gwhere-0.5B& \small Gwhere-1.5B&\small Gwhere-3B&\small Gwhere-7B\\
    \cmidrule{2-6}
    Acc@1 & 0.2675 & 0.2820 & 0.2862  & 0.2923 & 0.3071 \\
    % a-TCS & 0.5561 & 0.6071 & 0.6021 & 0.6632 & 0.7543 \\
    % a-SCS & 0.8022 & 0.8212 & 0.8231 & 0.8452 & 0.8838 \\
    % a-PAS & 0.6232 & 0.6285 & 0.6240 & 0.6531 & 0.6970 \\
    % a-SAS & 0.2472 & 0.2840 & 0.2988 & 0.3240 & 0.3725 \\
  \bottomrule
\end{tabular}
\end{table}

% \subsection{O}
\subsection{Overall Performance}
Our method substantially outperforms all baseline methods across all of the evaluated datasets. Table \ref{tab:nyc} illustrate the comparison results on NYC, TKY, and CA datasets. Specifically, compared to the state-of-the-art GNPR-SID which utilizes LLaMA-7B as its base model, we observe a substantial improvement on all of the versions of Gwhere. Notably, Gwhere-7B demonstrated the most significant improvements, achieving gains of 11.3\%, 8.1\%, and 13.2\% over the SOTA, respectively. 

Table \ref{tab:beijing} details the performance comparison between our proposed model and the cascade ranking system on the Amap dataset. Consistent with our findings on public datasets, all Gwhere variants significantly outperform the traditional cascade ranking system on Acc@1, with Gwhere-7B achieving 14.8\% relative improvement. The scaling trend from 0.5B to 7B confirms that larger model capacity benefits next POI prediction when combined with our SID-based generative paradigm.

\begin{table}[h!]
\centering
\caption{Comparison of different models on three datasets: NYC, TKY, and CA. We present the model inputs including the POI and User representation approach and whether visit timestamps are utilized. RID, SID, UID and UP represent Random one-hot ID, Semantic ID, User one-hot ID and User Profiles, respectively.  The results for the baseline methods are borrowed from \cite{zhang2024llm4poi, wang2025generative}.}
\label{tab:nyc}
\setlength{\tabcolsep}{4.pt}
\begin{tabular}{llcccccc}
\toprule
\multirow{2}{*}{Model} &  \multicolumn{3}{c}{Inputs} & & \multicolumn{3}{c}{Acc@1} \\
        \cmidrule{2-4} \cmidrule{6-8}
      & POI & User & Time  & & NYC & TKY & CA \\
\midrule
PRME    & RID & UID & $\times$ & & 0.1159 & 0.1052 & 0.0521 \\
% LSTM    & RID  & UID & $\times$ & & 0.1305 & 0.1335 & 0.0665 \\
PLSPL   & RID & UID & $\times$ & & 0.1917 & 0.1889 & 0.1072 \\
STAN    & RID  & UID & $\times$ & & 0.2231 & 0.1963 & 0.1104 \\
GETNext & RID & UID & $\times$ & & 0.2435 & 0.1829 & 0.1357 \\
STHGCN  & RID & UID & $\times$ & & 0.2734 & 0.2950 & 0.1730 \\
TPG     & RID  & UID & $\checkmark$ & & 0.2555 & 0.1420 & 0.1749 \\
ROTAN   & RID  & UID & $\checkmark$ & & 0.3106 & 0.2458 & 0.2199 \\
LLM4POI & RID & UID & $\checkmark$ & & 0.3372 & 0.3035 & 0.2065 \\
GNPR-SID & SID & UID & $\checkmark$ & & 0.3618 & 0.3062 & 0.2403 \\
\midrule
Gwhere-0.5B & SID & UP & $\checkmark$ & & 0.3705 & 0.3125 & 0.2555 \\
Gwhere-1.5B & SID & UP & $\checkmark$ & & 0.3727 & 0.3170 & 0.2585 \\
Gwhere-3B & SID & UP & $\checkmark$ & & 0.3872 & 0.3202 & 0.2611 \\
Gwhere-7B & SID & UP & $\checkmark$ & & \textbf{0.4027} & \textbf{0.3310} & \textbf{0.2721} \\
\bottomrule
\end{tabular}
% \caption*{\textit{The results for the base methods are borrowed from \cite{wang2025generative}.}}
\end{table}

\subsection{Ablation Study And Further Analysis}
\subsubsection{Ablation study}
To evaluate the impact of individual modules, we developed various versions of Gwhere. Initially, certain essential modules were removed: 
"\textbf{w/o SID}" replaces the proposed SID generation method with random id. 
"\textbf{w/o ST}" eliminates spatio-temporal descriptions from the prompt.
"\textbf{w/o SLD}" exclude the SID-Location-Description corpus in the continued pretraining stage.
"\textbf{w/o CPT}" exclude the Continued Pretraining stage in the post-training stage, only 
applying SFT and RL.
"\textbf{w/o SFT}" and "\textbf{w/o RL}" denote skipping SFT and RL training in the alignment stage, respectively. 
As illustrated in Table~\ref{tab:combined_left_right} Left, removing any module leads to noticeable Acc@1 degradation. In particular, removing SID tokenization or RL-based preference alignment leads to the largest performance drops, demonstrating their essential roles in recommendation effectiveness.
% The omission of CDI also causes a notable performance drop, affirming that our intervention steps are vital for sustaining high-quality training data.

% \begin{table}[h!]
%     \centering
%     \caption{Ablation study of Gwhere on AMAP Dataset(with the online deployment version Gwhere-3B).}
%     \label{tab:ablation}
%     \begin{tabular}{lccccc}
%         \toprule
%          Ablations & Acc@1  \\
%         \midrule
%         w/o SID & 0.2455 $(\downarrow 16.01\%)$  \\
%         % w/o UP & 0.2720 & 0.6435 & 0.8211 & 0.5920 & 0.2923  \\
%         w/o ST & 0.2785 $(\downarrow 4.72\%)$ \\
%         w/o SLD & 0.2822 $(\downarrow 3.46\%)$ \\
%         w/o CPT & 0.2632 $(\downarrow 9.96\%)$ \\
%         w/o SFT & 0.2725 $(\downarrow 6.77\%)$ \\
%         w/o RL & 0.2580 $(\downarrow 11.73\%)$ \\
%         % w/o DPO & 0.2710 & 0.6041 & 0.8028 & 0.6063 & 0.2785  \\
%         \midrule
%         Gwhere & 0.2923  \\
%         \bottomrule
%     \end{tabular}
% \end{table}

\subsubsection{Impact Analysis on Multi-modal Synergy}
\begin{figure}
    \centering
    \includegraphics[width=0.8\linewidth]{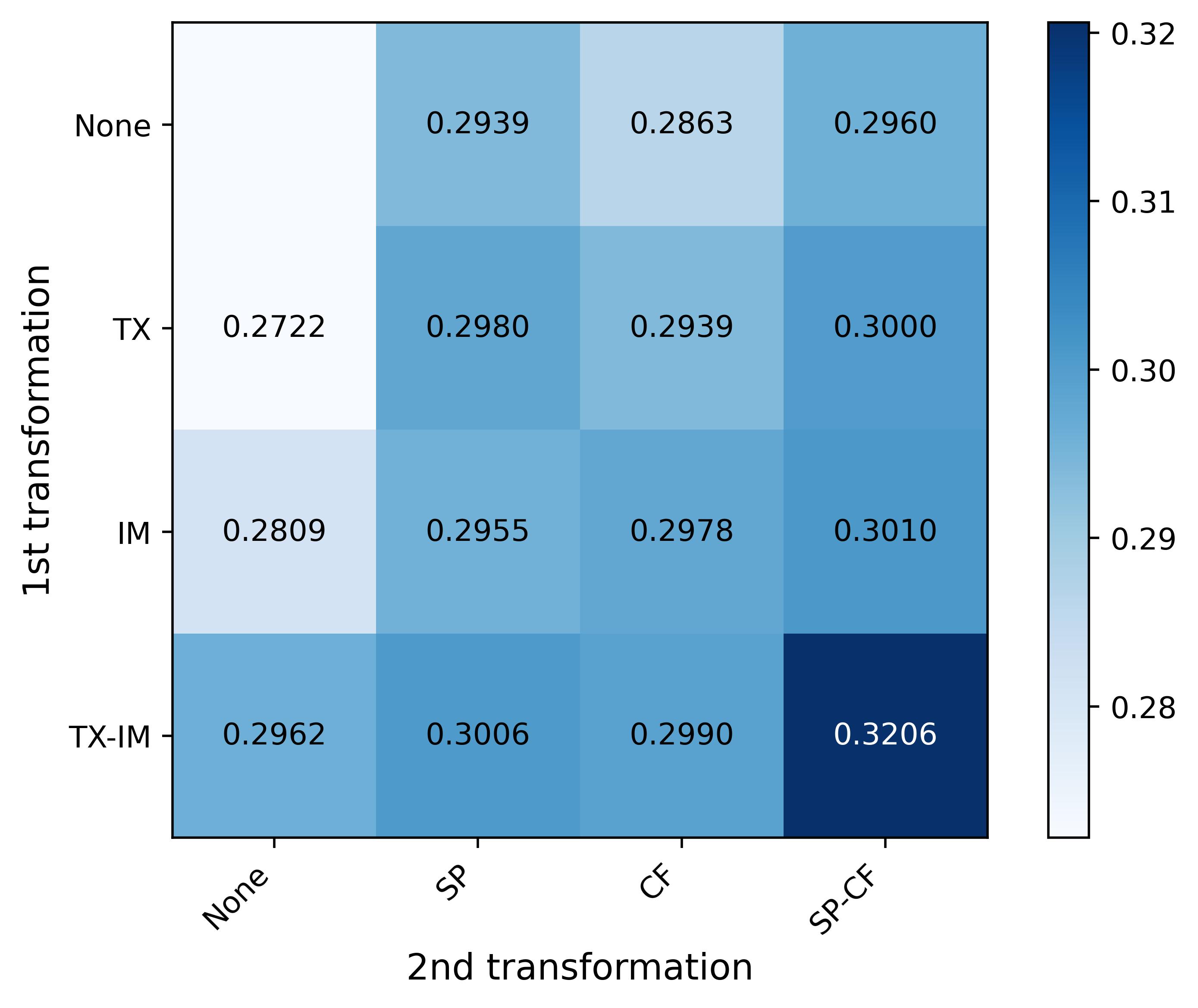}
    \captionsetup{aboveskip=3.5pt}
    \caption{Acc@1 results of AMAP under individual or composition of input modals. text is denoted as TX, IM is image, SP is spatial information and CF is collaborative signals. TX-IM (SP-CF) means composition of text-image (spatial-collaborative).}
    \label{fig:composition}
    \vspace{-1em}
\end{figure}

% SimCIT demonstrates native compatibility with heterogeneous item information sources through unified identifier alignment. 
We systematically evaluate modality-specific contributions using the AMAP dataset containing four distinct modalities: (1). textual descriptions (denoted as TX), (2). image content (denoted as IM), (3). Spatial information (denoted as SP) and (4). collaborative filtering signals (denoted as CF).
As shown in Fig. \ref{fig:composition}, progressive modality integration yields monotonically improving performance in generative recommendation tasks. Crucially, spatial features provide the largest individual gain. The full quad-modal configuration achieves state-of-the-art results, validating our framework's capacity to synthesize cross-modal synergies and preserve modality specificity.

% \subsubsection{Impact Analysis of Codebooks} The effectiveness of semantic tokenization can be influenced by key hyperparameters, such as the codebook size ($K$) the number of codebooks ($M$), and the embedding dimension ($d$) of code vectors. As illustrated in Fig. \ref{fig:abla} (Bottom), as the codebook size and the number of codebooks increases, NDCG
% shows a consistent upward trend. This suggests that a larger token space allows for more effective representation of each item, leading to more precise tokenization. Additionally, as the embedding dimension increases from 128 to 512, the model’s capability to represent items improves, leading to enhancements across various metrics. However, increasing the dimension to 256 results in a slight performance degradation due to overfitting.

\subsubsection{Impact Analysis on Preference Alignment}
To isolate the effect of the preference alignment strategy, we fix the SID tokenizer and model architecture and compare three variants: "SFT-only" is trained with continued pretraining plus SFT, but without any explicit preference optimization loss.
"SFT + DPO" denotes the baseline that adds a DPO-style loss over paired preferences derived from logged comparisons.
"SFT + GRPO" represents the baseline that utilizes GRPO strategy.
"SFT + EAKTO": our exposure-aware KTO setup using clicked POIs as desirable examples and exposed-but-unclicked POIs as E-Neg signals.

% \begin{table}[t]
%   \centering
%   \caption{Preference alignment on the AMAP next POI scenario. All models share the same SID tokenizer and backbone; only the preference objective differs. }
%   \label{tab:kto_ablation}
%   \begin{tabular}{lc}
%     \toprule
%     Variant & Acc@1  \\
%     \midrule
%     SFT-only & 0.2580 \\
%     SFT + DPO & 0.2690  \\
%     SFT + GRPO & 0.2716 \\
%     SFT + EAKTO & 0.2923 \\
%     \bottomrule
%   \end{tabular}
% \end{table}

Table~\ref{tab:combined_left_right} Right reports the comparison. EAKTO consistently outperforms DPO and GRPO, confirming its effectiveness in leveraging abundant exposure data under label imbalance.

\begin{table}[t!]
\centering
\caption{Left: Offline ablation study using Gwhere-3B on Amap dataset. Right: Preference alignment comparison on Amap.}
\label{tab:combined_left_right}
\resizebox{\linewidth}{!}{%
  \begin{tabular}{@{}cc@{}}
    % 左单元格：Gwhere 消融实验
    % \caption{Left: Ablation study of Gwhere on AMAP Dataset(with the online deployment version Gwhere-3B). Right: Preference alignment on the AMAP next POI scenario.}
    \begin{minipage}[c]{0.1\textwidth}
      \flushleft
      % \caption{Ablation study of Gwhere on AMAP Dataset(with the online deployment version Gwhere-3B).}
      \label{tab:ablation_left}
      \begin{tabular}{lc}
        \toprule
        \textbf{Ablations} & \textbf{Acc@1} \\
        \midrule
        w/o SID & 0.2455 $(\downarrow 16.01\%)$ \\
        w/o ST & 0.2785 $(\downarrow 4.72\%)$ \\
        w/o SLD & 0.2822 $(\downarrow 3.46\%)$ \\
        w/o CPT & 0.2632 $(\downarrow 9.96\%)$ \\
        w/o SFT & 0.2725 $(\downarrow 6.77\%)$ \\
        w/o RL & 0.2580 $(\downarrow 11.73\%)$ \\
        \midrule
        Gwhere & 0.2923 \\
        \bottomrule
      \end{tabular}
    \end{minipage} &
    % 右单元格：偏好对齐实验
    \begin{minipage}[c]{0.4\textwidth}
    % \caption{Preference alignment on the AMAP next POI scenario.}
      \centering
      \flushright
      \label{tab:preference_right}
      \begin{tabular}{lc}
        \toprule
        \textbf{Variants } & \textbf{Acc@1} \\
        \midrule
        SFT-only & 0.2492 \\
        SFT + DPO & 0.2634 \\
        SFT + GRPO & 0.2759 \\
        SFT + EAKTO & 0.2820 \\
        \bottomrule
      \end{tabular}
    \end{minipage} \\
  \end{tabular}%
}
\end{table}

\subsubsection{Impact Analysis on Training Corpus}
\begin{figure}[h!]
    \centering
    \includegraphics[width=\linewidth]{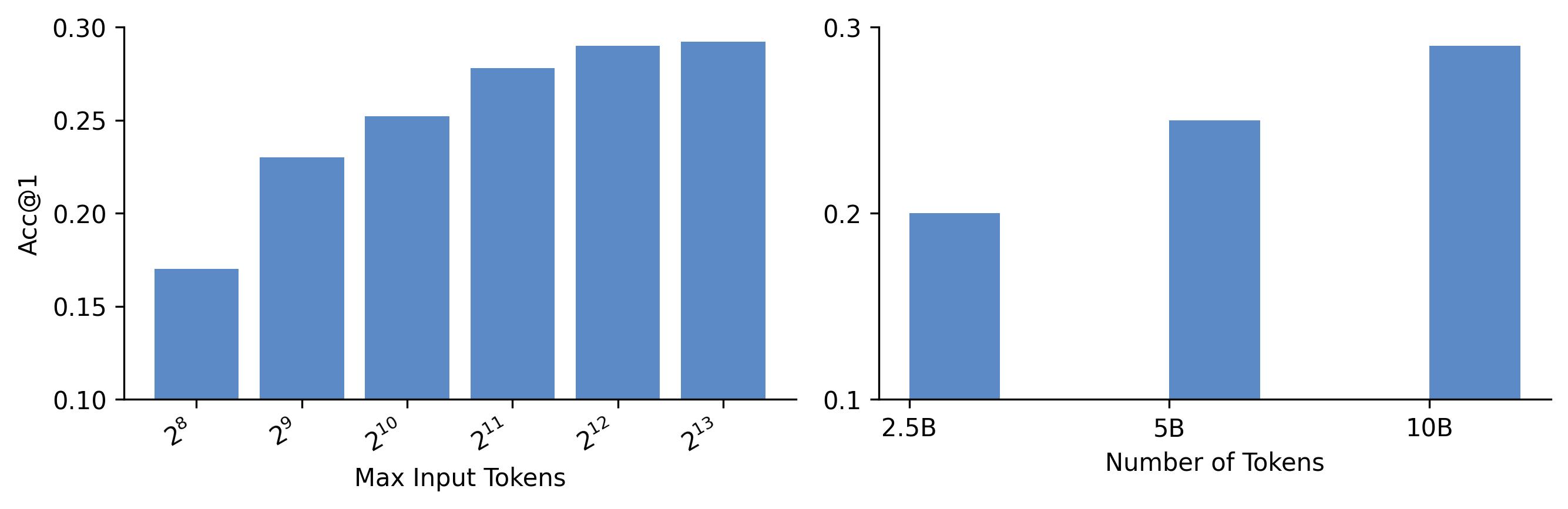}
    \captionsetup{aboveskip=3.5pt}
    \caption{Impact on the performance of sequence length (input tokens) and number of tokens involved in continued pretraining stage. }
    \label{fig:impact}
    \vspace{-1em}
\end{figure}

In this section, we analyze the impact on the performance of sequence length (input tokens) and number of utilized tokens in continued pretraining stage. Figure \ref{fig:impact} presents the results.
Our investigation into the impact of input sequence length shows that while longer context is beneficial, its utility eventually saturates. The model's performance gains are most pronounced up to a context window of 2048 tokens ($\sim$30 user historical check-ins) and largely stabilize beyond 4096 tokens ($\sim$50 check-ins). This suggests that a moderately-sized history is sufficient to encapsulate the user's core preferences. Further extending the context window provides limited new signal, as this long-term information is already implicitly encoded in the user's learned representation.

The volume of tokens involved in pretraining phase also proved to be a key determinant of performance. We observed a nearly linear improvement in model effectiveness as the training corpus size increased. Constrained by our training budget, the final model was trained to convergence on a dataset of 10B tokens. The consistent, non-saturating growth suggests that performance is largely data-bound, with significant potential for improvement as more data becomes available.

\subsection{Online Experiments}

\subsubsection{Scenario} Our method is performed on \textbf{Amap}, a prominent navigation and mapping platform hosting billions of users and POIs. We focus on the "\textbf{Guess Where You Go}" card presented on the homepage of the Amap App, aiming to forecast the next subsequent POI using a user’s previous check-ins and profile information, thereby improving key recommendation metrics, like CTR. In the long run, this work aims to enhance user engagement.

\paragraph{Production Baseline.}
The online baseline is the production cascade-ranking system used
by Amap for the same homepage entry. It consists of multiple
industrial recall channels, followed by coarse ranking, fine ranking,
and cognitive re-ranking. The system uses the same user request
context as Gwhere, including user profile, recent behavior, current
location, time, and situational signals, together with production
features such as exposure/click history, trending topics, and
seasonal patterns. Its optimization targets online engagement
metrics, including CTR-related objectives and user retention. Both
systems are evaluated under the same traffic allocation, page
position, request context, and metric definitions. Therefore, the
online A/B test compares Gwhere against the deployed production
baseline rather than a simplified offline ranking model.

% The objective of this scenario is to predict the next POI the users are most likely to visit. By displaying the POIs on the homepage, we aim to reduce the users' search effort, thereby enhancing user experience and long-term retention. 

\subsubsection{Efficiency Analysis}
\begin{figure}
    \centering
    \includegraphics[width=\linewidth]{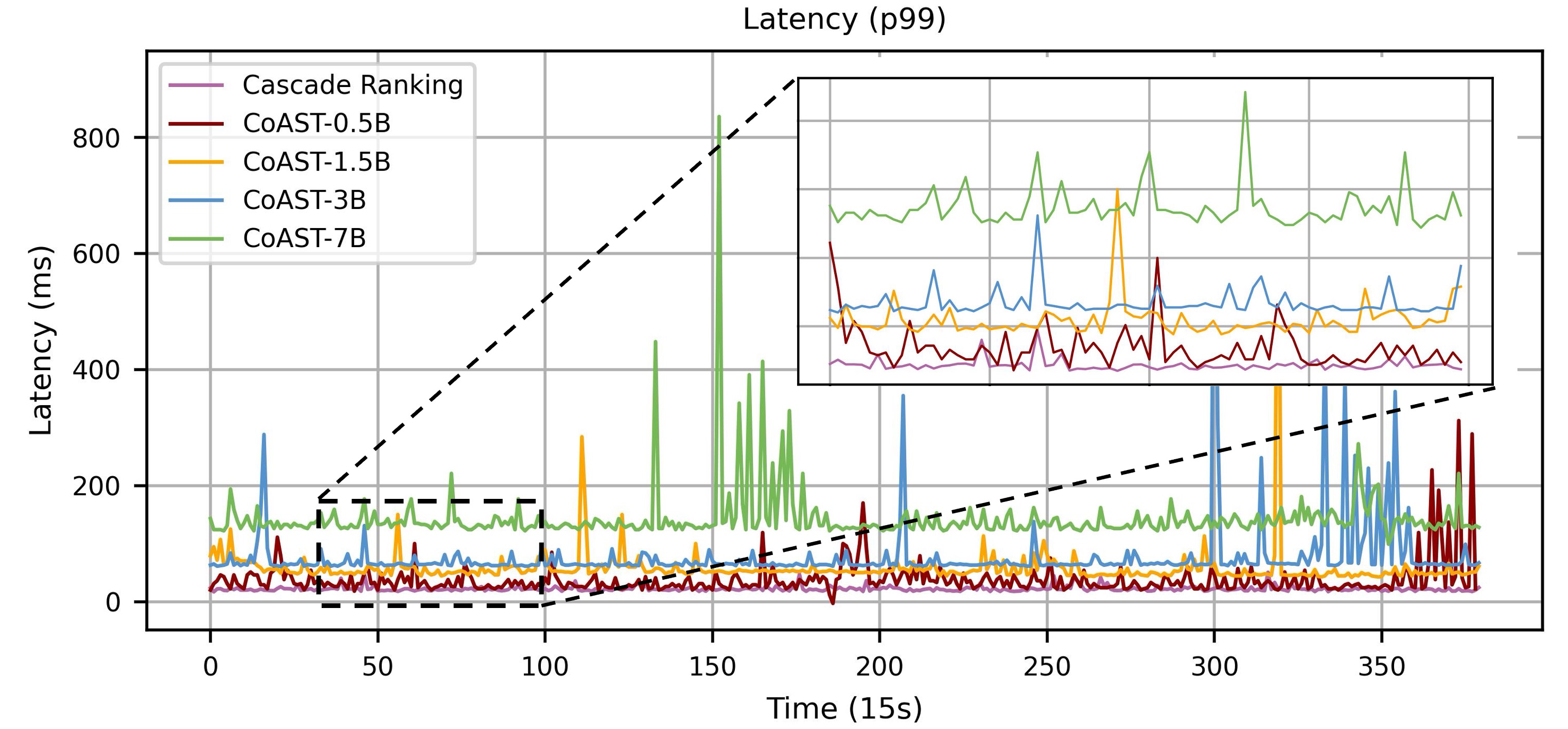}
    \captionsetup{aboveskip=5pt}
    \caption{Comparison of Model Inference p99 Latency (ms) under a 50 QPS Load Test on 2$\times$ NVIDIA H20 GPUs.}
    \label{fig:load-test}
    \vspace{-1em}
\end{figure}

To evaluate the online inference efficiency, we compare different variants of Gwhere with the production
cascade-ranking baseline described above. Before deploying our model, we conduct a load test under the settings with a 50 QPS on 2$\times$ NVIDIA H20 GPUs.
Figure \ref{fig:load-test} presents the inference p99 latency results of different methods. As shown, the Cascade Ranking system, a widely adopted paradigm in industrial recommender systems, demonstrates exceptional efficiency with its latency consistently around 20ms. As a comparison, while Gwhere-7B achieved the best offline performance among all our variants, its high computational cost posed a significant challenge for online serving. 

To strike a practical balance between efficacy and efficiency, we identified Gwhere-0.5B as the optimal candidate. It delivers competitive performance while maintaining a manageable inference cost ($\sim$ 30ms). Consequently, we deployed Gwhere-0.5B to production and conducted an online A/B test against the state-of-the-art (SOTA) Cascade Ranking baseline.

\subsubsection{Online A/B Test}
To assess the online effect of Gwhere, 
% with a focus on IPA,
we implemented a short-term and long-term A/B testing on "Guess Where You Go" card of Amap's homepage. We examined Gwhere-0.5B, trained based on a Qwen2.5-0.5B model, compared to the traditional cascade ranking baseline as shown in Table \ref{tab:online}. 
% \subsubsubsection{Short-term online A/B test} saf

\begin{table}[h]
    \centering
    \small 
    \setlength{\tabcolsep}{3pt}
    \caption{Online A/B test results on Amap's homepage service.The baseline is Amap's production cascade-ranking system described in Section~4.4.1. Results are reported as relative improvements (\%) over the production baseline. ``w/o EAKTO'' denotes the deployed Gwhere variant trained without exposure-aware preference optimization.
    }
    \label{tab:online}
    \begin{tabular}{lcc|ccc|c}
        \toprule
         &  P-CTR &  U-CTR &  SC-Rate & RE-Rate &   AC-Rate & NF-Rate \\
        \midrule
         w/o EAKTO &  +4.23\% &  +4.59\% &  +1.01\% &  +1.65\% &  +5.99\%  &  -5.85\% \\
         Gwhere &  +5.83\% &  +6.20\% &  +2.18\% &  +2.60\% &  +9.67\%  &  -11.11\% \\
        \bottomrule
    \end{tabular}
    \vspace{-0.5em}
\end{table}

It is worth mentioning that this baseline already includes user profiles, trending topics and seasonal features for improving cognition in the cognitive re-ranking stage.
% For U-CTR and CTR, percentage points (pt) were used to denote absolute differences. For instance, the / improvement in S-PVR for SerenGPT-IPO signifies a growth from / to /, equivalent to a / absolute rise.
For short-term (one week) online test, Gwhere demonstrates significant improvements in metrics like P-CTR (PV CTR) and U-CTR (UV CTR). This indicates that Gwhere has increased the share of cognitive-aligned recommendations that effectively attracted user attention in terms of clicks. 
Additionally, it slightly enhances the user engagement metrics, such as SC-Rate (Scroll Rate), RE-Rate (User Retention Rate), and AC-Rate (User Active Rate), implying heightened user involvement. 
Notably, the effects on general utility metrics such as CTR and NF-Rate (Users' Negative Feedback Rate) are more modest, while the improvements on these engagement-related metrics often become apparent only over the long term.

% The effects on general utility metrics such as CTR and number of total clicks ("U-" User View Rate and "P-" represents Page View Rate ) and total number of clicks are more modest. 
To investigate the long-term effectiveness of the proposed method in a large-scale industrial RS, we conducted an online A/B test over a one-month period. Compared to the traditional baseline, Gwhere achieves significant enhancements in user engagement metrics. Notably, it yields a 11.11\% decrease in NF-Rate (Negative Feedback Rate) and a 9.67\% increase in AC-Rate. These results demonstrate that Gwhere effectively mitigates irrelevant recommendations and improves user satisfaction, while simultaneously boosting CTR. Furthermore, the comparison with the ablation variant (w/o EAKTO) validates the critical role of our proposed module. As shown in Table~\ref{tab:online}, removing the EAKTO component leads to a notable performance drop, particularly in NF-Rate (from -11.11\% to -5.85\%). This significant gap confirms that the EAKTO mechanism is essential for balancing high user engagement with the effective suppression of negative feedback.

\subsection{Industrial Deployment and A/B Testing}
Deploying LLM-based generative retrieval in a large-scale industrial setting like Amap poses significant engineering challenges, primarily due to the ``token explosion'' phenomenon during auto-regressive generation. To ensure high concurrency and low latency, we designed a robust serving architecture with a strict closed-loop monitoring mechanism.

\textbf{Inference Optimization \& Monitoring:} We implemented continuous batching and PagedAttention to optimize KV-Cache memory. To guarantee system stability, we established a tiered monitoring system. The \textbf{P0 metrics} focus on generation bottlenecks, including Token Throughput (tokens/sec) and End-to-End (E2E) Latency. The \textbf{P1 metrics} track resource efficiency, such as KV-Cache hit rates and GPU memory fragmentation. Furthermore, we introduced an asynchronous handler mechanism with heartbeat checks to manage long-tail requests and prevent worker node timeouts during peak traffic.

\textbf{Online A/B Testing:} We deployed Gwhere in Amap's production environment, serving millions of active users. In a strict one-month online A/B test, Gwhere achieved a significant performance lift over the previous state-of-the-art production baseline, with a \(+5.83\%\) increase in Page Click-Through Rate (P-CTR) and a \(+6.20\%\) increase in User Click-Through Rate (U-CTR). These quantifiable business results demonstrate Gwhere's capability to drive core business conversions effectively.

\section{Conclusion}
\label{sec:conclusion}
In this work,
we propose a novel framework for next point-of-interest (POI) recommendation to address key limitations of traditional retrieve-rank systems, starting with semantic IDs (SIDs) generation via a contrastive learning-based tokenization mechanism that fuses multi-modal item information (textual descriptions, spatial features, visual content, and collaborative signals), moving beyond traditional reconstruction-based quantization to enhance token discriminability and alleviate code collision. We then adapt large language models (LLMs) to the mobility domain through continued pretraining on enriched spatio-temporal corpora, equipping them with domain-specific knowledge of geographical entities and sequential mobility patterns, followed by a multi-stage fine-tuning pipeline integrating Supervised Fine-Tuning (SFT) and reinforcement learning to implement a novel alignment strategy with real user behavior preferences. For industrial deployment, we adopt engineering optimizations including prefill-decoding decoupling and multi-token prediction, enabling the framework to meet high-concurrency and low-latency requirements in real-world location-based service (LBS) platforms. Extensive evaluations demonstrate that our framework outperforms traditional cascade ranking systems, and future work will focus on integrating more dynamic contextual signals, optimizing cold-start scenarios for new users and POIs, and exploring deeper alignment between semantic tokens and natural language to enhance recommendation interpretability for broader LBS applications.

% \begin{acks}
% We would like to acknowledge the discussions of Jiawei Xue, Ning Wang, and Fangfang Chen throughout this work. Furthermore, we are grateful for  Tucheng Lin, Jian Song, Shulong Han, Jinhui Chen, Xiongfei Fan, Zudan Cao, Jing sun, Yifei Fan and Yukun Liu, for their engineering support.
% \end{acks}

\bibliography{reference.bib}

\appendix

\end{document}